\documentclass[11pt,preprint]{aastex}

\newcommand{\etal}{{\it et~al.}}

\usepackage{longtable}
\begin{document}

\title{The Sensitivity of NEO Surveyor to Low-Perihelion Asteroids}

\author{Joseph R. Masiero\altaffilmark{1}, Yuna G. Kwon\altaffilmark{1}, Dar W. Dahlen\altaffilmark{1}, Frank J. Masci\altaffilmark{1},  Amy K. Mainzer\altaffilmark{2,3}}

\altaffiltext{1}{Caltech/IPAC, 1200 E. California Blvd, MC 100-22, Pasadena, CA 91125 USA}
\altaffiltext{2}{Department of Earth, Planetary and Space Sciences, University of California, Los Angeles, CA 90095 USA}
\altaffiltext{3}{University of Arizona, Tucson, AZ 85721 USA}

\begin{abstract}

Asteroids with low orbital perihelion distances experience extreme
heating from the Sun that can modify their surfaces and trigger
non-typical activity mechanisms.  These objects are generally
difficult to observe from ground-based telescopes due to their
frequent proximity to the Sun.  The Near Earth Object Surveyor
mission, however, will regularly survey down to Solar elongations of
$45^\circ$ and is well-suited for the detection and
characterization of low-perihelion asteroids.  Here, we use the survey
simulation software tools developed for mission verification to
explore the expected sensitivity of NEO Surveyor to these objects.  We
find that NEO Surveyor is expected to be $>90\%$ complete for near-Sun
objects larger than $D\sim300~$m.  Additionally, if the asteroid
(3200) Phaethon underwent a disruption event in the past to form the
Geminid meteor stream, Surveyor will be $>90\%$ complete to any
fragments larger than $D\sim200~$m.  For probable disruption models,
NEO Surveyor would be expected to detect dozens of objects on
Phaethon-like orbits, compared to a predicted background population of
only a handful of asteroids, setting strong constraints on the
likelihood of this scenario.

\end{abstract}

\section{Introduction}

Asteroids, as reservoirs of materials from the early inner Solar
system \citep{demeo15}, allow us to study the mineralogical processes
that occurred in the early protosolar disk.  The main asteroid belt
between Mars and Jupiter is generally stable over the age of the Solar
system, meaning most objects found there have been largely unmodified
since their formation \citep{binzel15}. However, some of these objects
can drift into resonances that change their orbits so that they
approach close to the Earth.  These objects, with perihelia less than
$q<1.3~$AU, are known as near-Earth objects (NEOs) and experience very
different evolutionary regimes for their $\sim10~$Myr dynamical
lifetime \citep{gladman00}.

A very small subset of NEOs evolve onto orbits with extremely low
perihelion distances.  Currently, only 28 of the $>33,000$ known
near-Earth asteroids, along with comet 96P/Machholz, have perihelia of
$q<0.15~$AU.\footnote{We do not consider sun-grazing comets; while the
SOHO mission has discovered well over 1000, they are generally thought
to have a different formation history (originating mostly from the
Oort Cloud as opposed to the inner solar system), and many of them are
destroyed shortly after discovery.}  These objects experience subsolar
heating that can be in excess of $1,000~$K \citep{maclennan21},
exposing primitive surface materials to these temperature regimes for
the first time.  The evolution of the surface materials in this
regime, as well as their potential for catastrophic disruption
\citep[cf.][]{granvik16}, provides important clues that help us
investigate the global physical properties of these objects.

The poster-child for low-perihelion asteroids is (3200) Phaethon,
which will soon be visited by the DESTINY+ mission \citep{arai23}.
Discovered during the infrared sky survey conducted by IRAS
\citep{green85a}, Phaethon was assumed to be an extinct cometary
nucleus due to its orbital proximity to the Geminid meteor stream
\citep{green85b}.  Although numerous studies of this object were
undertaken, detection of any activity was not seen until a small tail
was identified by \citet{jewitt13} in STEREO data during the object's
2009 and 2012 perihelion passes.  Multiple theories have been put
forth for the driver of observed activity, including thermal
fracturing \citep{ryabova18} and volatilization of sodium
\citep{masiero21,zhang23}, but all are a result of the extreme
temperatures Phaethon experiences at perihelion.

Phaethon's association with the Geminid stream has led to the
suggestion that a breakup event thousands of years ago might explain
the association between these objects and other nearby asteroids like
(155140) 2005 UD \citep{ohtsuka06}.  Given that the currently observed
levels of activity are insufficient to populate the Geminid stream
\citep{jewitt10}, such a breakup event could naturally resolve this
discrepancy.  \citet{devogele20} showed that the physical properties
of Phaethon and 2005 UD are consistent, strengthening the proposed
link, while dynamical modeling by \citet{jo24} finds an optimal epoch
of dust creation $\sim18~$kyr ago.  A fission event large enough to
create the Geminid stream and 2005 UD from a proto-Phaethon body would
also be expected to produce a number of intermediate sized objects in
the tens-to-hundreds of meter range that would still be present in the
NEO population today, based on observed outcomes of rotational
\citep{jewitt17} and tidal \citep{sekanina94} breakups.

The NEO Surveyor mission \citep{mainzer23} will conduct a census of
asteroids and comets near the Earth's orbit, in order to determine the
risk posed to our planet from any potential impactors.  The mission
makes use of a single instrument consisting of a two-channel thermal
infrared camera, and will conduct a dedicated survey optimized for the
detection of near-Earth objects.  NEO Surveyor is expected to increase
the catalog of known NEOs by more than an order of magnitude,
detecting two-thirds of all potentially hazardous asteroids larger
than $140~$m in diameter after 5 years and $90\%$ after 12 years.  As
part of its survey, NEO Surveyor will regularly observe down to Solar
elongations of $45^\circ$ covering a region of space where
low-perihelion asteroids spend a significant fraction of time.  In
this work, we investigate the potential of NEO Surveyor for
discovering and characterizing low-perihelion asteroids, and we use these
simulations to give predictions on the number of Phaethon-like objects
that will be detected.  We can also set constraints on the mission's
sensitivity to any asteroids created if Phaethon previously underwent
a breakup event.

\section{Population Model}

Our work makes use of the recently developed survey simulation tools
for the NEO Surveyor mission \citep{mainzer23}.  These tools have been
demonstrated by \citet{masiero23} to successfully reproduce predicted
observations of near-Earth asteroids through comparison with data from
the NEOWISE mission \citep{mainzer11,mainzer14neowise,mainzer19}.
Validation of the predicted positions, fluxes, and detectabilities
ensures that survey simulation outputs match the expectations for
performance of the mission system in flight.

To build our input population, we take all known asteroids with
perihelia less than $q<0.15~$AU that have detections over multiple
orbital epochs from the Minor Planet Center (MPC) orbit
catalog\footnote{\it https://minorplanetcenter.net/iau/MPCORB.html}.
At the time this work was carried out this subset included 28 objects.
Although a relatively small sample, the distribution of orbital
elements for these objects showed:
\begin{itemize}
\item Perihelion distance $q$ was consistent with a flat distribution from $\sim0.075~$AU to the cutoff at $0.15~$AU,
\item Eccentricity $e$ is approximately a Gaussian that peaks at
  $\sim0.92$ and is truncated at $0.97~$AU (with two outliers at
  $0.7<e<0.8$),
\item Inclination $i$ is roughly flat below $\sim35^\circ$ (with one
  outlier above $i>50^\circ$),
\item The argument of perihelion and longitude of the ascending node are consistent with flat distributions across all angles,
\item No significant correlations exist between $q-e$, $q-i$, or $e-i$.
\end{itemize}

\begin{figure}[ht]
\begin{center}
  \includegraphics[scale=0.7]{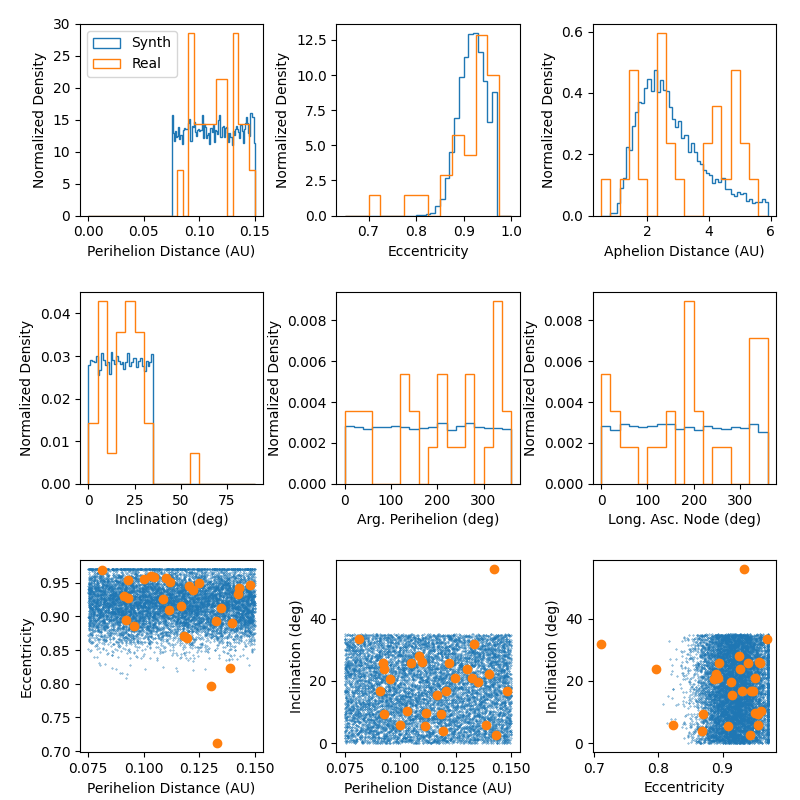}
  \protect\caption{Comparison of the currently known population of
    near-Sun asteroids (orange) to the synthetic population used for
    our completeness determinations (blue).  Histograms of orbital
    parameters (top left: perihelion distance; top middle:
    eccentricity; top right: aphelion distance; middle left: orbital
    inclination; center: argument of perihelion; middle right:
    longitude of the ascending node) and scatter plots of orbital
    elements (bottom left: perihelion distance vs eccentricity; bottom
    middle: perihelion distance vs inclination; bottom right:
    eccentricity vs inclination) demonstrate that the synthesized
    population sufficiently samples the phase space occupied by the
    real objects.}
\label{fig.population}
\end{center}
\end{figure}

In order to perform a statistically significant simulation, we take
these observed trends and synthesize a population of 10,000 objects to
run through our survey simulator.  Time of perihelion for each object
is chosen from a flat distribution within 4000 days of the start of
survey (the approximate period of our most extreme object with
$q=0.15$ and $e=0.97$). Figure \ref{fig.population} shows the real
objects from the MPC catalog in orange, and the synthetic population
in blue.  The synthetic population is broadly consistent with the real
objects (neglecting the few outliers), and sufficient to allow us to
estimate the sensitivity of NEO Surveyor to unknown near-Sun objects.

\clearpage

Our choice of synthetic orbital parameters restricts our input
population to having aphelion outside of Q$\sim0.7~$AU.  This is
coincident with the known population, however the known objects are
almost certainly biased against low aphelia by the difficulty in
observing these objects from ground- or space-based facilities.  Due
to the design of the spacecraft sunshade, NEO Surveyor is restricted
to observing objects with heliocentric distances larger than
$R_h>0.7~$AU, and thus objects with Q$<0.7~$AU can never be detected.
Although a population of such objects may well be present in the inner
Solar system, it neither poses a hazard to Earth nor will be
observable by NEO Surveyor, and so we do not consider it further in
this analysis.

For tests of sensitivity to objects that might have been formed from a
breakup event of a proto-Phaethon object, we generate a more focused
synthetic population of 1,000 objects around Phaethon's current
orbital parameters to better test our sensitivity to this region of
phase space.  Here we use flat distributions in a narrow range around
Phaethon's orbit: $q=0.14\pm0.01$, $e=0.890\pm0.015$,
$i=22\pm5$, $0\le\omega\le360$, and $0\le\Omega\le360$.  Time of
perihelion is chosen from a flat distribution with $523~$days of the
survey start, consistent with the period of Phaethon.  This population
of Phaethon-like objects provided improved statistics for sensitivity
as a function of diameter, as discussed below.

\section{Survey Simulation Results}

We use the NEO Surveyor Survey Simulator (NSS) software tools
\citep{mainzer23} to propagate our synthetic orbits over the 5 year
nominal mission survey lifetime and calculate their expected fluxes
with respect to the instrument sensitivity.  The NSS then determines
if the objects are detectable a sufficient number of times to report
tracklets (sequences of position-time measurements) to the MPC that
would all for the orbit to be constrained.  State vector propagation
is carried out using an N-body simulator; instrument sensitivity uses
the Current Best Estimate for the hardware that will be implemented
for the mission; tracklet generation efficiency is based on on-going
tests of the NEO Surveyor Science Data System at IPAC; estimations for
the capability of the MPC to link tracklets into measurable orbits are
based on historic performance from the NEOWISE mission as well as
on-going testing between NEO Surveyor and the MPC. \citet{mainzer23}
provide the specific values used for these parameters in the NSS and
the method of their derivation.  The output of the NSS is a list of
observations of each input object that passes detection and
tracklet-building requirements; this is used to determine overall
completeness as a function of both size and orbit, described in detail
below.

\subsection{Size Effects}

\begin{figure}[ht]
\begin{center}
  \includegraphics[scale=0.7]{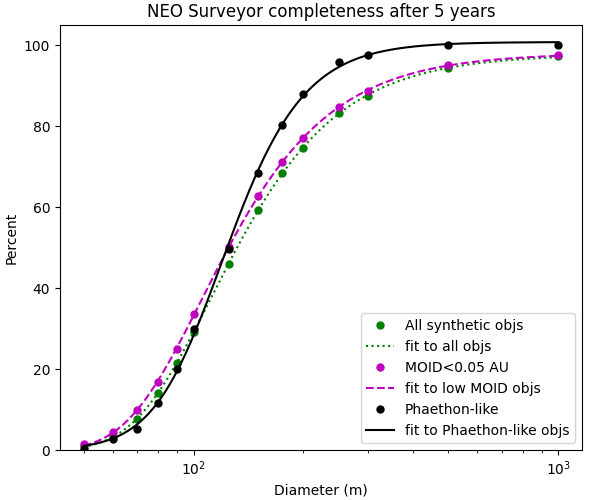}
  \protect\caption{Fraction of synthetic objects recovered by NEO
    Surveyor after 5 years as a function of object size.  Green points
    show all synthetic objects, magenta shows those objects with Earth
    MOIDs less that 0.05 AU, and black shows a subset of objects close
    in orbital space to (3200) Phaethon.  The lines for each color
    show the best fit to the calculated completeness points using
    Eq~\ref{eq.compfit}. }
\label{fig.diameter}
\end{center}
\end{figure}

The larger an object is, the more flux it will emit for a given
orbital position and observing geometry, and so the easier it will be
to detect.  Since larger objects have more opportunities to be
detected, it also increases the chance that they will be cataloged
following our survey rules.  In this analysis we take our synthetic
orbital population, assign all objects the same diameter, and run the
survey simulator to determine completeness at the end of five years.
By sweeping through a range of diameters we can constrain the
completeness as a function of size.  We show these results in
Figure~\ref{fig.diameter}.

The general behavior of the full synthetic population shows that while
$50~$m objects will have low completeness fractions, this rises
rapidly with size and surpasses $90\%$ for sizes of a few hundred
meters.  Restricting our analysis to the synthetic objects that come
closest to Earth's orbit, those with Earth Minimum Orbit Intersection
Distance (MOID) less then $0.05~$AU, we see that the completeness
fraction is generally a few percent higher at all sizes until reaching
the same saturation point at a few hundred meters.  Objects with
Phaethon like orbits follow a comparable trend, though have a steeper
change in completeness with size and reach $>90\%$ completeness at
smaller sizes of $D\sim200~$m.  This improvement compared to the
general synthetic population is primarily due to the relatively lower
eccentricity of the Phaethon-like set of objects which increases the
overall likelihood of detection.

We can analytically describe the completeness vs size by fitting a
generalized logistic function to each of our results, of the form:
\begin{equation}
  F = L  \left( 1+e^{-k(s-s_0)} \right) ^\beta
\label{eq.compfit}
\end{equation}
\noindent where F is the completeness fraction, L is the maximum
fraction, s is the $\log$ of the size (in meters), s$_0$ is the pivot
point of the function (in $log(m)$), k is the steepness of growth, and
$\beta$ dictates the asymmetry of the growth.  The best fit values
for each parameter are given in Table~\ref{tab.fits}.  We note that
the best-fit maximum fraction for Phaethon-like objects settles to
slightly above $100\%$; this is due to the incomplete sampling of the
size range at very large sizes. Artificially constraining this to an
upper limit of $100\%$ produced a worse fit to the data, and so we
choose to leave it as is, with the caveat that it is simply a fitting
artifact.

\begin{table}[ht]
\begin{center}
\scriptsize
  \noindent
  \caption{Parameters of the best-fit logistic functions to the survey completeness}
  \vspace{1ex}
  {
  \noindent
  \begin{tabular}{ccccc}
  \tableline
Synthetic Population & k & s$_0$ (log(m)) & $\beta$ & L \\
  \tableline

All near-sun &  5.19 &   1.60 & -10.08 &  97.8\\
Low MOID     &  5.12 &   1.51 & -13.97 &  98.1\\
Phaethon-like & 8.50 &   2.02 &  -1.67 & 100.8\\

\hline
  \end{tabular}
  }
  \label{tab.fits}
\end{center}
\end{table}

\subsection{Orbital Element Dependencies}

For objects in near-Earth space, geometry can conspire to cause some
classes of orbits to only approach Earth's orbit when the planet is far
from the close-approach point.  NEO Surveyor will have improved sensitivity
to these objects through searching the sky at low Solar elongations,
scanning down to $45^\circ$ ahead and behind the Earth
along its orbit.  Even with this survey strategy, however, some very high
eccentricity asteroids can require many years of survey before falling
into one of these fields of regard.  

This is particularly true for low-perihelion objects with high
eccentricities which can spend large fractions of their orbit too
close to the Sun to be observable or too far from the Sun to be
detectable.  Surveyor's five-year nominal mission survey provides
opportunities to sample the full orbit for objects with semimajor axes
$a<3~$AU, increasing the likelihood of detection.  The majority of our
synthetic sample is within this orbital range, though longer survey
periods would be expected to improve completeness by recovering
objects on near-resonant orbits with Earth's orbit.

\begin{figure}[ht]
\begin{center}
  \includegraphics[scale=0.5]{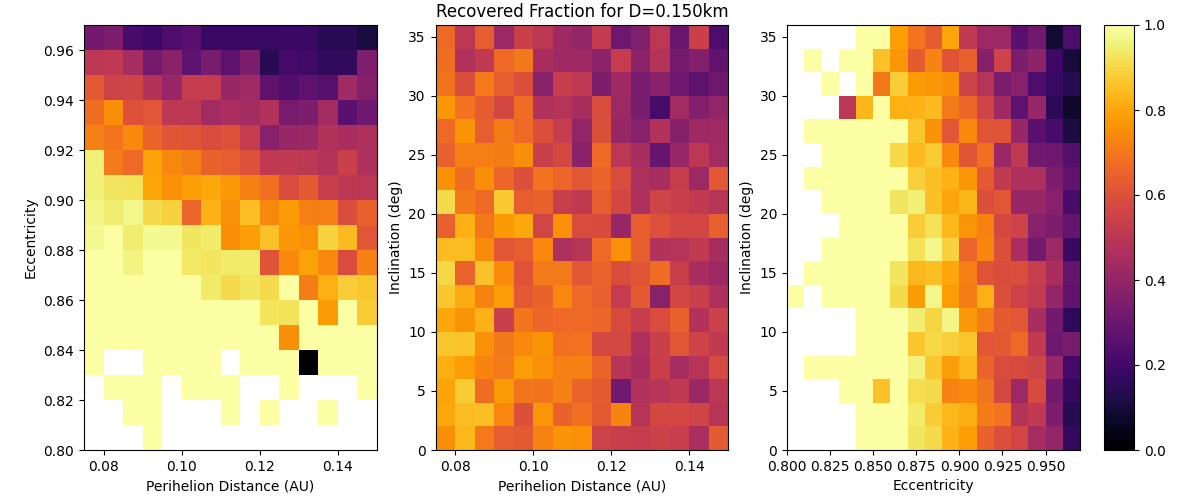}
  \protect\caption{Fraction of synthetic objects recovered by NEO
    Surveyor after 5 years (colorbar) as a function of orbital
    elements assuming all objects have diameters of D$=150~$m.
    Eccentricity vs perihelion distance (left) shows a strong
    correlation with recovery fraction; inclination vs perihelion
    distance (middle) shows only a weak trend; inclination vs
    eccentricity (right) shows that eccentricity dominates the
    recoverability. }
\label{fig.elements}
\end{center}
\end{figure}

In Figure~\ref{fig.elements} we show the fraction of objects that
would be detected and cataloged by NEO Surveyor after five years as a
function of their orbital elements.  Bins showing eccentricities of
$e<0.85$ have few objects in them based on our input population, and
in some cases are not shown where no objects were implanted. A
semimajor axis of 1$~$AU approximately traces the edge of the highest
completeness bins in the plot of perihelion distance vs eccentricity
in Figure~\ref{fig.elements}.

We see that overall, NEO Surveyor will have very high completeness for
most of our synthetic population, with low completenesses seen only
for the highest eccentricity and highest perihelion distance objects.
This class of objects benefits most from longer duration surveys, and
so would be expected to grow in completeness as survey continues
toward the twelve-year mission goal.  Orbital inclination does not
show a significant effect on completeness, at least in the parameter
range probed by our synthetic population.  For different object
diameters, completeness in each bin follows the same trend as
discussed above, growing with size until completeness reaches $100\%$.
This completeness `front' tends to follow lines of constant semimajor
axis in the perihelion-eccentricity plot.

\section{Discussion}

Phaethon's current level of activity is insufficient to sustain the
Geminid meteor population that we observe today, however it has been
postulated to have undergone a major breakup event in the last million
years that would have created the Phaethon-Geminid Complex including
(potentially) the asteroids (155140) 2005 UD and (225416) 1999 YC
\citep{ohtsuka06,ohtsuka09}.  If such an event happened, we would
expect the population of objects created to follow a size-frequency
distribution similar to other observed breakup events, leaving the
$6.12$-by-$4.14~$km diameter Phaethon as the largest remnant
\citep{yoshida23}.

\begin{figure}[ht]
\begin{center}
  \includegraphics[scale=1]{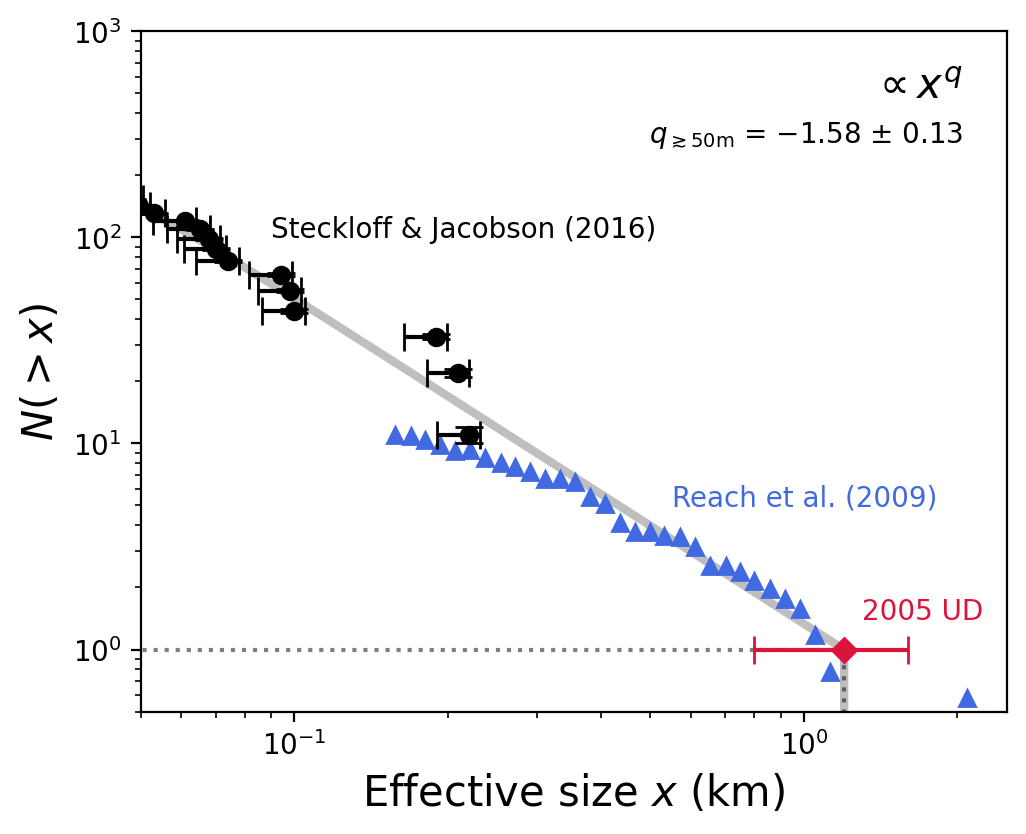}
  \protect\caption{Cumulative size frequency distribution ($q$, grey line)
    of expected breakup distribution resulting in 2005 UD as the
    largest fragment (red diamond).  Over-plotted are cometary fragment
    distributions determined by \citet{reach09} (blue triangles) and
    \citet{steckloff16} (black circles) for analogous situations.}
\label{fig.sfd}
\end{center}
\end{figure}

Using 2005 UD as the largest fragment \citep[with a diameter of
  $1.2~$km][]{masiero19}, we constructed a cumulative size frequency
distribution (SFD) of cometary fragments larger than $\sim50~$m in
diameter. The result with a power-law index $q$ is shown in
Figure~\ref{fig.sfd}. The best estimate and error indicate the median
and standard deviation from the literature, respectively. A trend line
in this size regime is determined by using three literature studies
that provide approximate size ranges: (1) An SFD of stria-forming
chunks from Comet West, likely fragmented by water-ice sublimation
from \citet{steckloff16}; (2) an SFD of major fragments of comet
73P/Schwassmann-Wachmann 3 observed by Spitzer Space Telescope from
\citet{reach09}, whose upper-limit size is estimated from the flux at
24 $\mu$m using an approximate equation; and (3) an SFD of fragments
of C/1999 S4 (LINEAR) observed by SWAN telescope from
\citet{makinen01}. The trend lines and data points in shown are scaled
arbitrarily once 2005 UD is set at (size=1.2 km, N=1) as a hard
reference point, enabling the high-confidence portion of the data
set's size regime to align with the trend line. Based on the
estimates, we anticipate that there will be $\sim$140 near-Earth
objects besides 2005 UD with a diameter between $\sim$50 m and
$\sim$1.2 km.

\begin{figure}[ht]
\begin{center}
  \includegraphics[scale=0.8]{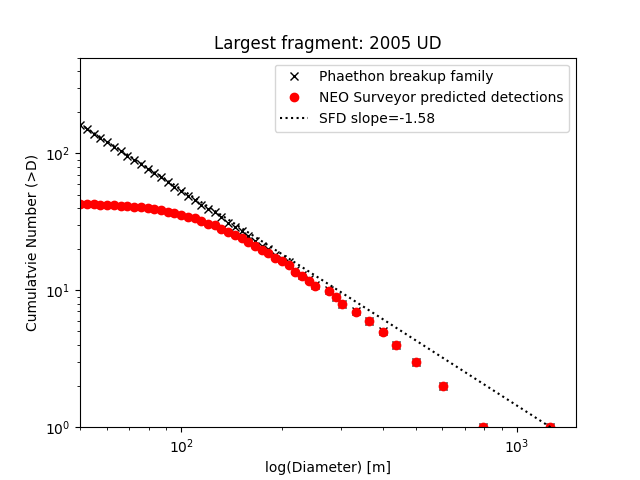}
  \protect\caption{Cumulative size frequency distribution of the
    objects present today in the hypothesized Phaethon-Geminid Complex
    formed from the breakup of a proto-Phaethon object (black x's) and
    the subpopulation that would be expected to be detected and
    cataloged by NEO Surveyor (red circles), under the assumption that
    the largest fragment of the breakup is (155140) 2005 UD.  NEO
    Surveyor will be approximately complete to D$\sim200~$m, and would
    detect $\sim40$ members of this population.}
\label{fig.found}
\end{center}
\end{figure}

We take this size frequency distribution, and apply our survey
completeness parameters for objects on Phaethon-like orbits
(Table~\ref{tab.fits}) to determine the expected number of objects
that NEO Surveyor will detect from the hypothesized Phaethon-Geminid
Complex, which we show in Figure~\ref{fig.found}.  Following the
diameter-completeness plots above, the five-year survey carried out by
NEO Surveyor would be effectively complete to $\sim200~$m for this
population.

In total, $43$ objects from a breakup event would be expected to be
detected over the course of the 5-year survey.  This is a
significantly larger than the $3$ objects that would be expected to be
detected in this orbital element phase space from the population
present in the NEO Surveyor Reference Small Body Population Model
\citep[RSBPM][]{mainzer23}.  Currently in this region of orbital
element space there is only $1$ object known besides Phaethon: 2007
PR10, which is an insufficient sample to differentiate a breakup
complex from the background population.  NEO Surveyor will thus
provide sufficient data to confirm or refute the existence of a
breakup-created Phaethon-Geminid Complex.  In the event that our
assumption that 2005 UD is the largest breakup fragment is incorrect,
and instead a smaller object such as the $\sim600~$m (504181) 2006 TC
is the largest remnant, NEO Surveyor would still be expected to detect
$13$ objects over the 5-year survey in this region of orbital element
phase space, a statistically significant excess above the background.

\section{Conclusions}

Using the new software tools developed to validate mission
requirements for NEO Surveyor, we have studied the mission's
sensitivity to near-Sun asteroids, including objects with orbital
elements similar to (3200) Phaethon.  We find that NEO Surveyor is
highly sensitive to this population, with $50\%$ completeness reached
for objects of size D$\sim125~$m after 5 years and $90\%$ completeness
for objects with D$>400~$m for a wide range of near-Sun orbits.
Objects on Phaethon-like orbits would be expected to have
completenesses of $90\%$ for sizes of D$>200~$m.  If the Geminid
meteor stream formed from a breakup event of a proto-Phaethon object,
NEO Surveyor would detect and catalog dozens of members of such a
group, allowing us to test this formation theory.  This contrasts with
the handful of background NEOs in this orbital element space that
would be expected to be detected from the nominal RSBPM used for
survey evaluation.  The data that will be obtained from NEO Surveyor
will significantly improve our understanding of objects on near-Sun
orbits.

\section*{Acknowledgments}
We thank the two anonymous referees for their careful review of this
work.  This publication makes use of software and data products from
the NEO Surveyor, which is a joint project of the University of
Arizona and the Jet Propulsion Laboratory/California Institute of
Technology, funded by the National Aeronautics and Space
Administration.  This research has made use of data and services
provided by the International Astronomical Union's Minor Planet
Center.  This research has made use of NASA’s Astrophysics Data System
Bibliographic Services.  This research has made use of the {\it
  numpy}, {\it scipy}, {\it astropy}, and {\it matplotlib} Python
packages.

\end{document}